\documentclass[twocolumn]{aastex63}

\usepackage{xspace, comment, amsmath}
\usepackage{float}
\newcommand{\snf}{SNfactory\xspace}
\newcommand{\stdstar}{standard-star\xspace}

\newcommand{\um}{\ensuremath{\mu}m\xspace}
\newcommand{\AAA}{\AA\xspace}

\submitjournal{PASP}

\shorttitle{The SCAT Survey}
\shortauthors{M. A. Tucker et al.}
\graphicspath{{./}{figures/}}
\begin{document}

\title{The Spectroscopic Classification of Astronomical Transients (SCAT) Survey: \\ Overview, Pipeline Description, Initial Results, and Future Plans}

\correspondingauthor{Michael Tucker}
\email{tuckerma95@gmail.com}

\author[0000-0002-2471-8442]{M. A. Tucker}
\altaffiliation{CCAPP Fellow}
\affiliation{Center for Cosmology and Astroparticle Physics,
The Ohio State University, 
191 West Woodruff Ave,
Columbus, OH, USA}
\affiliation{Department of Astronomy, 
The Ohio State University, 
140 West 18th Avenue,
Columbus, OH, USA}
\affiliation{Department of Physics,
The Ohio State University,
191 West Woodruff Ave,
Columbus, OH, USA}

\author{B. J. Shappee}
\affiliation{Institute for Astronomy,
University of Hawai`i at M\={a}noa,
2680 Woodlawn Dr., Honolulu, HI, USA}

\author{M. E. Huber}
\affiliation{Institute for Astronomy,
University of Hawai`i at M\={a}noa,
2680 Woodlawn Dr., Honolulu, HI, USA}

\author{A. V. Payne}
\affiliation{Institute for Astronomy,
University of Hawai`i at M\={a}noa,
2680 Woodlawn Dr., Honolulu, HI, USA}

\author{A. Do}
\affiliation{Institute for Astronomy,
University of Hawai`i at M\={a}noa,
2680 Woodlawn Dr., Honolulu, HI, USA}

\author{J. T. Hinkle}
\affiliation{Institute for Astronomy,
University of Hawai`i at M\={a}noa,
2680 Woodlawn Dr., Honolulu, HI, USA}

\author[0000-0001-6069-1139]{T. de Jaeger}
\affiliation{Institute for Astronomy,
University of Hawai`i at M\={a}noa,
2680 Woodlawn Dr., Honolulu, HI, USA}

\author{C. Ashall}
\affiliation{Department of Physics, Virginia Tech, Blacksburg, VA 24061, USA}

\author[0000-0002-2164-859X]{D. D. Desai}
\affiliation{Institute for Astronomy,
University of Hawai`i at M\={a}noa,
2680 Woodlawn Dr., Honolulu, HI, USA}

\author[0000-0003-3953-9532]{W. B. Hoogendam}
\altaffiliation{NSF Graduate Research Fellow}
\affiliation{Institute for Astronomy,
University of Hawai`i at M\={a}noa,
2680 Woodlawn Dr., Honolulu, HI, USA}

\author{G. Aldering}
\affiliation{Lawrence Berkeley National Laboratory, One Cyclotron Rd., Berkeley, CA, 94720, USA}

\author[0000-0002-4449-9152]{K.~Auchettl}
\affiliation{School of Physics, The University of Melbourne, Parkville, VIC 3010, Australia}
\affiliation{ARC Centre of Excellence for All Sky Astrophysics in 3 Dimensions (ASTRO 3D)}
\affiliation{Department of Astronomy and Astrophysics, University of California, Santa Cruz, CA 95064, USA}

\author[0000-0002-1917-9157]{C. Baranec}
\affiliation{Institute for Astronomy, University of Hawai`i at M\={a}noa, Hilo, HI 96720, USA}

\author{J. Bulger}
\affiliation{Institute for Astronomy,
University of Hawai`i at M\={a}noa,
2680 Woodlawn Dr., Honolulu, HI, USA}

\author[0000-0001-6965-7789]{K. Chambers}
\affiliation{Institute for Astronomy,
University of Hawai`i at M\={a}noa,
2680 Woodlawn Dr., Honolulu, HI, USA}

\author[0000-0002-8462-0703]{M. Chun}
\affiliation{Institute for Astronomy, University of Hawai`i at M\={a}noa, Hilo, HI 96720, USA}

\author{K. W. Hodapp}
\affiliation{Institute for Astronomy, University of Hawai`i at M\={a}noa, Hilo, HI 96720, USA}

\author{T. B. Lowe}
\affiliation{Institute for Astronomy,
University of Hawai`i at M\={a}noa,
2680 Woodlawn Dr., Honolulu, HI, USA}

\author{L. McKay}
\affiliation{Institute for Astronomy, University of Hawai`i at M\={a}noa, Hilo, HI 96720, USA}

\author{R. Rampy}
\affiliation{Institute for Astronomy,
University of Hawai`i at M\={a}noa,
2680 Woodlawn Dr., Honolulu, HI, USA}

\author{D. Rubin}
\affiliation{Institute for Astronomy,
University of Hawai`i at M\={a}noa,
2680 Woodlawn Dr., Honolulu, HI, USA}

\author{J. L. Tonry}
\affiliation{Institute for Astronomy,
University of Hawai`i at M\={a}noa,
2680 Woodlawn Dr., Honolulu, HI, USA}



\begin{abstract}

We present the Spectroscopic Classification of Astronomical Transients (SCAT) survey, which is dedicated to spectrophotometric observations of transient objects such as supernovae and tidal disruption events. SCAT uses the SuperNova Integral-Field Spectrograph (SNIFS) on the University of Hawai'i 2.2-meter (UH2.2m) telescope. SNIFS was designed specifically for accurate transient spectrophotometry, including absolute flux calibration and host-galaxy removal. We describe the data reduction and calibration pipeline including spectral extraction, telluric correction, atmospheric characterization, nightly photometricity, and spectrophotometric precision. We achieve $\lesssim 5\%$ spectrophotometry  across the full optical wavelength range ($3500-9000$~\AAA) under photometric conditions. The inclusion of photometry from the SNIFS multi-filter mosaic imager allows for decent spectrophotometric calibration ($10-20\%$) even under unfavorable weather/atmospheric conditions. SCAT obtained $\approx 640$ spectra of transients over the first 3 years of operations, including supernovae of all types, active galactic nuclei, cataclysmic variables, and rare transients such as superluminous supernovae and tidal disruption events. These observations will provide the community with benchmark spectrophotometry to constrain the next generation of hydrodynamic and radiative transfer models. 

\end{abstract}



\section{Introduction}\label{sec:intro}

The past decade in astronomy has seen a huge increase in the amount of observational data available to the community, largely due to the proliferation of imaging surveys. Some surveys, such as the All-Sky Automated Search for SuperNovae \citep[ASAS-SN; ][]{shappee2014, kochanek2017} and the Asteroid Terrestrial-impact Last Alert System \citep[ATLAS; ][]{tonry2018}, cover the entire sky at moderate ($\lesssim 19$~mag) depth. Complementary surveys, such as the Panoramic Survey Telescope and Rapid Response System \citep[Pan-STARRS; ]{chambers2016}, the Zwicky Transient Facility \citep[ZTF; ][]{bellm2019}, and the upcoming Large Synoptic Survey Telescope \citep[LSST; ][]{ivezic2019}, provide or will provide deeper coverage for select portions of the sky.

Nearly every field of astronomy has benefited from the calibrated time-series photometry supplied by such surveys. However, these surveys have been particularly beneficial for studying astrophysical transients. These objects, such as supernovae (SNe) and tidal disruption events (TDEs), are stochastic events that provide fleeting glimpses into the physical processes governing extreme events in the Universe. In the past, a few hundred SNe were discovered per year by targeted galaxy surveys \citep[e.g., ][]{hamuy1993, li2000, aldering2002}, but the current era of imaging surveys has increased the transient discovery rate hundred-fold \citep[e.g., ][]{kulkarni2020}. This deluge of discoveries has revealed both new types of transients, such as ``fast blue optical transients'' \citep[FBOTs; e.g., ][]{prentice2018, ho2021}, as well as extending our knowledge of rare and unique versions of known events, such as the faint SNe Iax class \citep[e.g., ][]{woodvasey2002, li2003, foley2013} of thermonuclear explosions. 

While imaging surveys have undoubtedly transformed our understanding of the Universe and the myriad of explosive events that can occur within it, even well-sampled multi-filter photometry has its limitations. Spectroscopy is necessary for accurate transient classification and for constraining physical quantities such as temperature, density, and velocity which are essential for constructing a useful model. In addition to the program described in this manuscript, other notable transient classification programs include the ZTF Bright Transient Survey (BTS; \citealp{fremling2020, perley2020}) and the ``advanced'' extended Public ESO Spectroscopic Survey of Transient Objects (ePESSTO+, \citealp{smartt2015}). Time-series spectroscopy, where multiple spectra are obtained over days or weeks, provides a view into how these quantities evolve with time, which, in turn, probes the underlying powering mechanism(s) \citep[e.g., ][]{ashall2014, nicholl2014, holoien2020, dimitriadis2022}. 

However, most spectroscopic observations are obtained with slit spectrographs, where a finite-width slit is placed before the dispersive element in the optical path. The slit ensures stable spectral resolution and prevents contamination from other sources in the field of view (FoV) but makes absolute flux calibration\footnote{When referencing absolute fluxes in this work, we are implicitly referring to the CALSPEC system \citep{bohlin2014} which provides absolute flux scalings accurate to $\approx 1\%$.} difficult. The amount of light passing through the slit depends on the slit alignment and the atmospheric seeing. Additionally, atmospheric differential refraction (ADR) introduces a wavelength dependence on the amount of light passing through the slit \citep[e.g., ][]{filippenko1982}, especially for long exposures. This complicates the comparison between observations and physical models.

Spectrophotometry, where dispersed spectra are calibrated to an absolute flux scale, is \textit{possible} for slit spectrographs in some cases. Time-series spectra combined with well-sampled high-quality photometry can produce decent spectrophotometric results by ``mangling'' or ``warping'' the observed spectra to match the photometry \citep[e.g., ][]{hsiao2007}. However, this process is imperfect, especially for objects that do not have smooth spectral energy distributions due to strong absorption or emission features. 

Integral field units (IFUs) or integral field spectrographs (IFSs) simultaneously disperse an entire FoV which is ideal for spectrophotometry of transients. Slit effects are no longer a concern, retaining absolute photometry capabilities, and observations of spectrophotometric standard stars ensure a reliable spectral shape. Additionally, most transients occur in external galaxies and require mitigation of coincident host-galaxy light. The combined spectral and spatial information provided by IFU observations is essential for accurate host-galaxy removal and, therefore, absolute calibration of the transient spectrum. 

Here we describe the Spectroscopic Classification of Astronomical Transients (SCAT) survey which is dedicated to spectrophotometric observations of transient phenomena. \S\ref{sec:facilities} describes the survey motivation and facilities. Then, we outline the data reduction and calibration procedures in \S\ref{sec:pipeline}. \S\ref{sec:spectrophotometry} describes the necessary steps for placing the observed spectra on a reliable absolute flux scale. Finally, preliminary results and future developments are presented in \S\ref{sec:conclusion}. 

\section{Survey Overview}\label{sec:facilities}

\subsection{Motivation}

The SCAT survey is designed to maximize the scientific return of imaging sky surveys. The science objectives are two-fold: publicly classify bright, nearby, and/or interesting transients, and initiate spectrophotometric follow-up for the most intriguing objects.

Classification targets are primarily selected based on brightness, and we impose a limiting magnitude of $\approx 19$~mag (in optical filters) which provides a decent-quality extracted spectrum (S/N $\sim 10$) in $\approx 30$~minutes under typical conditions (quarter moon, thin/cirrus clouds). After extracting the spectrum with a custom quick-reduction pipeline, all transient classifications are released to the public via the Transient Name Server (TNS)\footnote{\url{https://www.wis-tns.org}}. We caution that the quick-reduction pipeline does not ensure absolute flux calibration (see \S\ref{subsec:discuss.classify}).

Spectrophotometric follow-up observations are obtained for particularly interesting transient sources. Follow-up targets generally fall into two categories: bright or unique. Bright transients are ideal for obtaining well-sampled high-quality spectrophotometry with the purpose of testing our understanding of the physical processes involved. Unique transients offer a different view into transient phenomena, as these sources are often poorly characterized and any data is useful to the community as we strive to constrain the mechanism(s) governing their formation and evolution. A subset of recent results using SCAT data is described in \S\ref{sec:conclusion}.

\subsection{Telescope and Instrumentation}

The SCAT survey utilizes the SuperNova Integral Field Spectrograph \citep[SNIFS; ][]{lantz2004} on the University of Hawai'i 2.2~m (UH2.2m) telescope. SNIFS was built by the Nearby Supernova Factory \citep[SNfactory; ][]{aldering2002}, and expressly designed to deliver transient spectrophotometry. SNIFS is comprised of three components: the photometric P channel and the two spectroscopic (B+R) channels. 

\begin{figure}
    \centering
    \includegraphics[width=\linewidth]{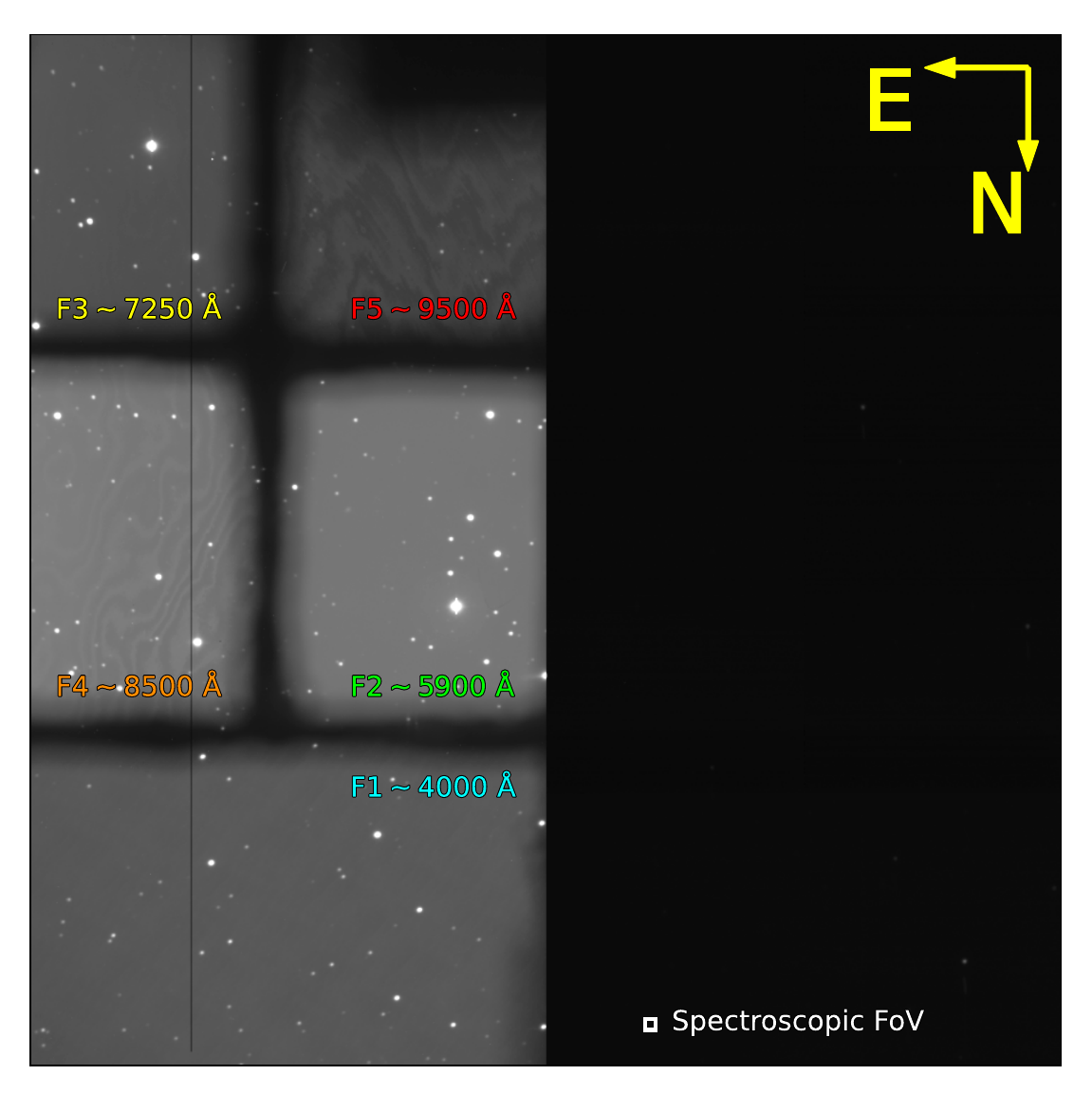}
    \caption{Example image of the multi-filter (MF) mosaic with effective wavelengths marked for each filter. The lower-right white square represents the $6\arcsec\times6\arcsec$ FoV of the spectroscopic channels.}
    \label{fig:mfexample}
\end{figure}

The P channel is responsible for acquiring, imaging, and guiding. The current filter set includes SDSS $ugriz$ plus Bessel $V$-band and the multi-filter (MF) mosaic, described below. Each image covers $9\arcmin\times9\arcmin$ with a pixel scale of $0\farcs137$~per pixel. The full-frame image is comprised of 2 CCDs with a 77-pixel ($\approx 10\arcsec$) vertical chip gap in the center. Under typical conditions, a $20$~s $V$-band acquisition image has a limiting magnitude of $V\approx 19$~mag. A unique component of the SNIFS P channel is the inclusion of the MF imaging setup comprised of 5 custom filters shown in Fig.~\ref{fig:mfexample}. The filters are designed to probe different aspects of the atmospheric throughput \citep{buton2009}.

Light is sent to the spectroscopic B and R channels using a pick-off prism (POP). After deflection by the POP, incident light is split between the B and R channels with a dichroic mirror then sent to the microlens arrays, each having a $6\arcsec\times6\arcsec$ FoV. The B channel covers $\approx 3400-5100$\AAA and the R channel covers $\approx 5100-10000$~\AAA, providing complete spectroscopic coverage across the full optical range. The spectral resolutions of the B and R channels are $\approx 5$~\AAA and $\approx 7$~\AAA, respectively. 

\vspace{0.7cm}

\section{SCAT Data Reduction and Calibration}\label{sec:pipeline}

The performance of the SNIFS+UH2.2m optical system has been extensively tested and calibrated by the SNfactory team \citep{aldering2002}. SNfactory is dedicated to time-series spectrophotometry of Type Ia supernovae (SNe Ia) for measuring cosmological parameters. To achieve this goal, they have calibrated the SNIFS spectrophotometric response to within $0.3\%$ of the Hubble Space Telescope (HST) CALSPEC system \citep{rubin2022}, the de facto ``gold standard'' in spectrophotometry \citep{bohlin2014}. SNfactory is able to reach this level of precision by dedicating a significant amount of time and effort to calibrating each aspect of the optical system, including extensive \stdstar observations \citep[e.g., ][]{buton2013} and the implementation of the SNIFS Calibration Apparatus \citep[SCALA; ][]{scala1, scala2, scala3}. Additionally, the SNfactory data reduction pipeline includes several algorithms that are not implemented in our pipeline. These corrections, such as the binary offset effect in CCD images \citep{boone2018}, mitigate sources of noise below our desired threshold.

As outlined in \S\ref{sec:facilities}, our goal is distinct from that of the \snf team. SCAT observes transients for understanding the physical processes governing the transient phenomena that occur throughout the Universe. We prioritize integration time on science targets by obtaining minimal calibration data throughout the night. Here we detail the photometric data reduction process in \S\ref{subsec:pipeline.P} and the spectroscopic procedure in \S\ref{subsec:pipeline.BR}. This section draws extensively from \citet{pereira2008} and \citet{buton2009} which laid the groundwork for the absolute calibration of SNIFS observations. 

\subsection{P Channel}\label{subsec:pipeline.P}

\begin{figure}
    \centering
    \includegraphics[width=\linewidth]{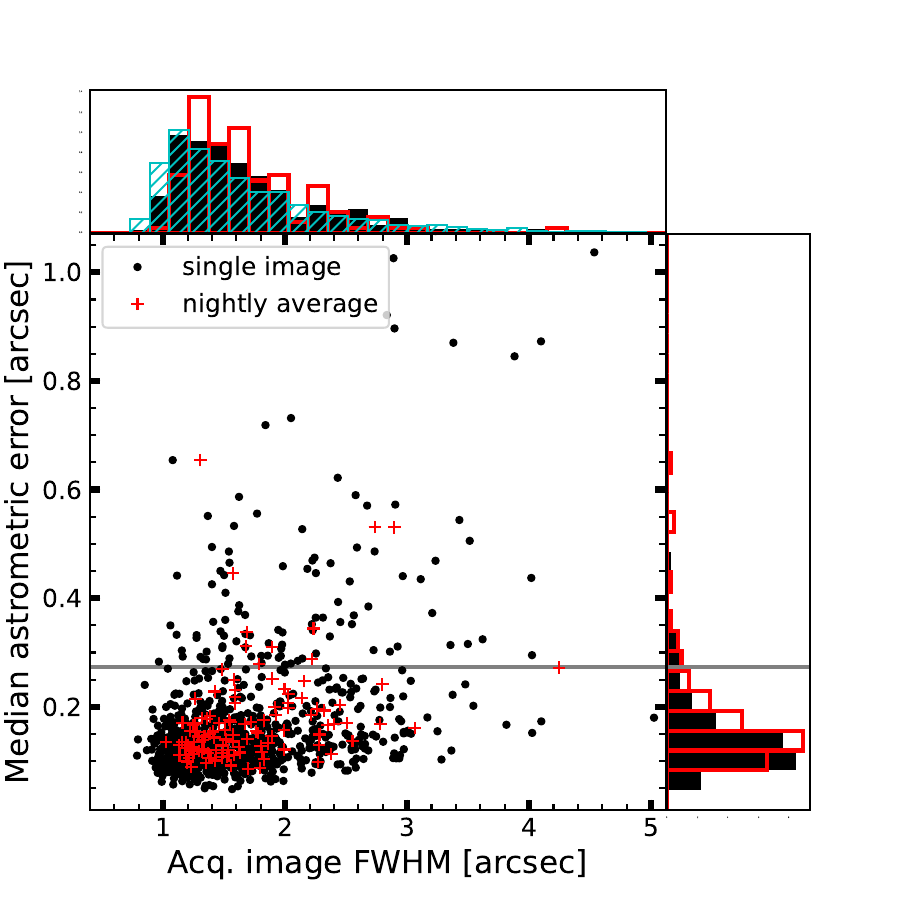}
    \caption{Comparison of the acquisition image FWHM to the median astrometric error. Black and red points represent single-exposure and nightly averaged values, respectively. The blue hatched histogram along the top shows the guider FWHM measurements from Fig.~\ref{fig:guider}. Acquisition images are obtained with $2\times2$ binning resulting in a pixel scale of $0\farcs27$/pixel (grey horizontal line).}
    \label{fig:acqfwhm}
\end{figure}

\begin{figure}
    \centering
    \includegraphics[width=\linewidth]{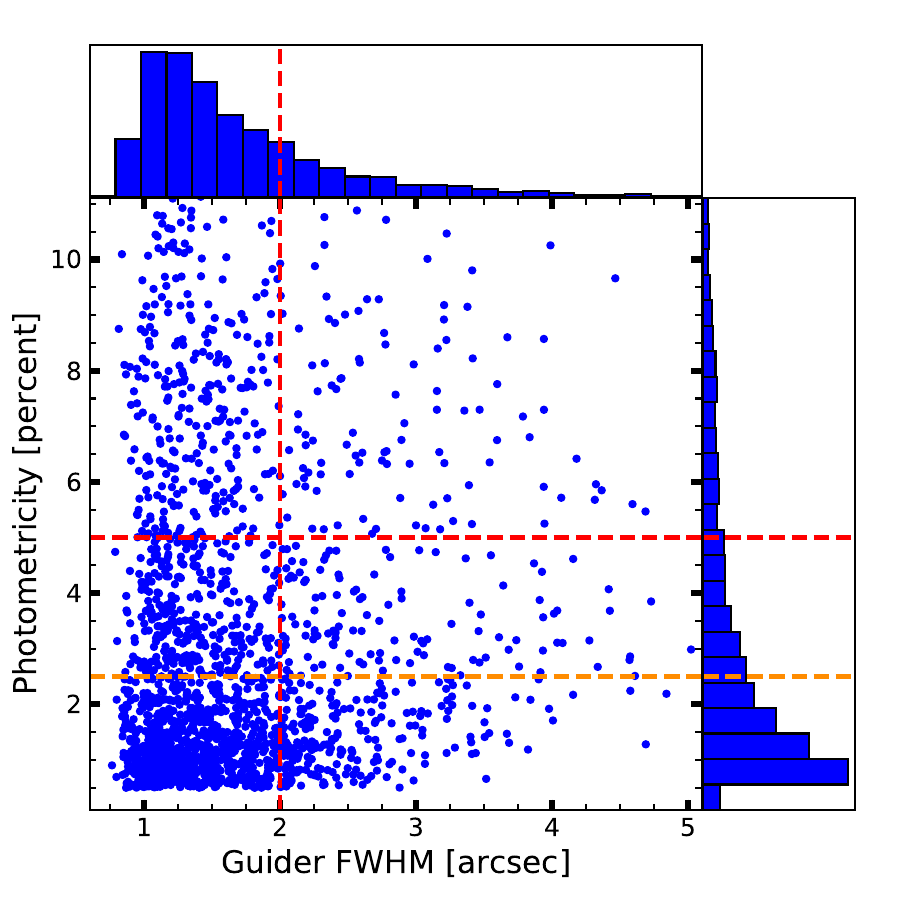}
    \caption{Comparison of the guider FWHM to the photometricity. Dashed lines represent the thresholds for photometric (orange) and non-photometric (red) conditions (see \S\ref{subsec:specphot.photometricity}).}
    \label{fig:guider}
\end{figure}

The P-channel images are overscan and dark subtracted, then flat-fielded using a median-combined sky flat constructed from 200 randomly-selected exposures taken during adjacent observing nights, with a maximum allowable time difference of 2 weeks. Exposures with short integration times ($< 5~\rm s$) are excluded to avoid bright ($V\sim 6$~mag) standard stars. Finally, the astrometric solution is derived using the \textsc{astrometry.net} software package \citep{lang2010}. Fig.~\ref{fig:acqfwhm} compares the median astrometric error, measured as the offset between the reference coordinates and the coordinates derived from the source positions, to the image FWHM. The majority of acquisition images have sub-pixel astrometric precision.  

During spectroscopic exposures, stars that fall in the SNIFS P-channel FoV are imaged concurrently through one of the available filters, usually the MF or $V$-band filter. The MF images cannot be astrometrically calibrated with the \textsc{astrometry.net} software due to the inter-filter gaps and the low number of reference stars. Instead, we use the telescope offset between the $V$-band acquisition image and the MF image to obtain a coarse alignment which is then tweaked (up to a few arc-seconds) by centroiding stars detected in both frames.

After astrometric calibration, we perform PSF-fitting photometry on each image. The full-frame $V$-band acquisition images have enough stars where the PSF can be derived independently for each image. For the MF images, we apply the wavelength-dependent PSF fit from the spectroscopic exposure (see \S\ref{subsec:pipeline.BR}) because the spectroscopic and photometric PSFs are similar \citep{pereira2008}. Finally, we zeropoint each image using the \textsc{refcat} catalog \citep{refcatref}. Stars used to model the PSF are required to have no sources within 5\arcsec and $12 < m ~\rm{[mag]} < 18$. The brightness and crowding limitations are removed when applying the PSF to sources detected in the image except for the exclusion of saturated objects. Sources that are separated by $< 2 \times $FWHM have their PSF fluxes fit simultaneously with a common background level.

\subsection{B and R channels}\label{subsec:pipeline.BR}

\begin{figure}
    \centering
    \includegraphics[width=\linewidth]{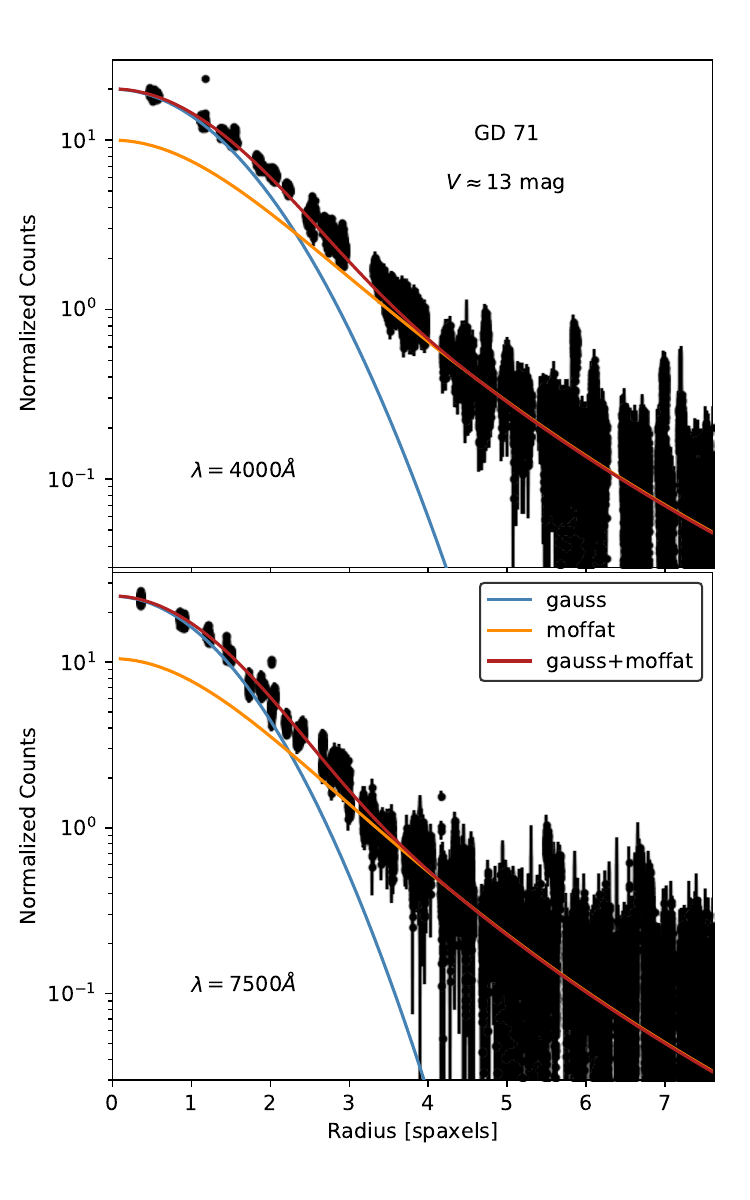}
    \caption{Example PSF fits to the B (top) and R (bottom) channels for a faint spectrophotometric standard star, GD71, including the Gaussian core (blue), the Moffat wings (orange), and the total PSF (red).}
    \label{fig:specpsf}
\end{figure}

The overall spectroscopic reduction process is similar for both B and R channels. The 2D spectroscopic exposures are pre-processed by the UH2.2m summit computers including bias and overscan subtraction, bad pixel masking, wavelength calibration with arc-lamp exposures, and finally converted into 3D $(x,y,\lambda)$ datacubes \citep{bacon2001}. Then, the extracted datacubes are transferred to our data reduction server. The datacubes are flat-fielded with nightly continuum lamp exposures and cosmic-ray hits are detected and interpolated over. Then, the extraction trace and ellipticity are estimated by fitting a wavelength-dependent 2D Gaussian profile to ``meta-slices'' with widths of $100~$\AAA and $150~$\AAA for the B and R channels, respectively. Then, we construct the normalized radial PSF which is well-described by a Gaussian core and Moffat wings \citep{buton2009, rubin2022};

\begin{equation}\label{eq:psf}
    PSF ( r_e, \sigma, \eta, \alpha, \gamma) = \eta \times G(r_e, \sigma) + M(r_e, \alpha, \gamma)
\end{equation}

\noindent where $G(r)$ describes the Gaussian core with standard deviation $\sigma$, $M(r)$ represents the Moffat wings with radial scale term $\gamma$ and power-law index $\alpha$, and $\eta$ determines the contribution from each component. The PSF may not be exactly circular, usually due to small guiding errors and slight atmospheric distortions, so we compute the elliptical radius 

\begin{equation}
    r_e^2 = (x - x_0)^2 + A(y-y_0)^2 + 2B(x-x_0)(y-y_0)
\end{equation}
\noindent where $(x_0, y_0)$ represent the center of the PSF and the ellipticity is described by the parameters $A$ and $B$. 

\begin{figure}
    \centering
    \includegraphics[width=\linewidth]{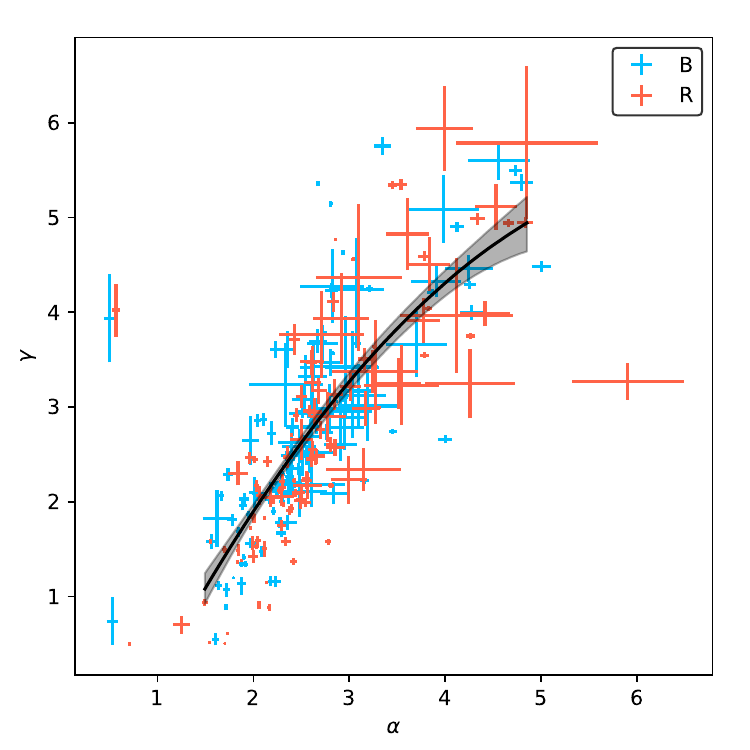}
    \caption{Correlation between the Moffat power-law index $\alpha$ and the radial term $\gamma$ for the central wavelengths of the B and R channels. The black line represents a quadratic fit to the combined data with the shaded region signifying the uncertainty. Fitting each channel separately produces a correlation in agreement with the fit to the combined data to better than $0.5\sigma$.}
    \label{fig:alphagamma}
\end{figure}

Standard-star spectra are extracted by fitting 

\begin{equation}\label{eq:spextract}
    F_\lambda = I_\lambda \times PSF(r_e, \sigma, \eta, \alpha, \gamma) + c_\lambda
\end{equation}

\noindent to the 3D cube using a Markov Chain Monte Carlo (MCMC) sampler with a $\chi^2$ minimization loss function and uninformative priors on the parameters. The observed flux $F_\lambda$ is comprised of the incident source flux, $I_\lambda$, modulated by the PSF (Eq. \ref{eq:psf}) and the addition of a spatially uniform background $c_\lambda$. We build a PSF template library using the \stdstar spectra with ``good" PSF fits (i.e., $\chi^2/\nu \sim 1$) to measure parameter correlations and wavelength dependencies. 

These parameter correlations are used as informative priors when extracting science exposures. Fig.~\ref{fig:alphagamma} shows that the Moffat parameters $\alpha$ and $\gamma$ can be tied during the fitting process and we use the fitted correlation (the black line in Fig.~\ref{fig:alphagamma}) as a Gaussian prior at each $\lambda$. Additionally, the Gaussian FWHM is set to the guide star FWHM at $\lambda \approx 5500~\rm \AA$ (i.e., $V$-band). This reduces the number of variables in the PSF fit to just 2, and therefore only 4 free parameters in Eq.~\ref{eq:spextract}: $I_\lambda$, $c_\lambda$, $\eta$, and a joint parameter representing the Moffat $\alpha$ and $\gamma$ terms. 

\section{Spectrophotometry and Absolute Flux Calibration}\label{sec:spectrophotometry}

\begin{table*}[]
    \centering
    \begin{tabular}{lcc}
        
        \hline\hline
        Probe & Photometric & Non-photometric \\
        \hline
        \multicolumn{3}{c}{\underline{Single Exposure Criteria}} \\
        PSF fit $\chi^2/\nu$ & $\leq1.2$ & $>1.5$ \\
        Guide star flux RMS & $\leq2.5\%$ & $> 5\%$ \\
        \multicolumn{3}{c}{\underline{Nightly Criteria}} \\
        SkyProbe Transmission RMS & $\leq2.5\%$ & $>5\%$ \\
        Median Seeing & $<1.5\arcsec$ & $> 2\arcsec$\\
        $V$-band zeropoint uncert. & $\leq 0.03$~mag & $>0.05$~mag \\
        MF zeropoint uncert. & $\leq 0.05$~mag & $>0.1$~mag \\
        Standard-star flux solution RMS & $\leq3\%$ & $>5\%$  \\
        Photometric exposure fraction & 1 & $< 0.75$ \\
        \hline\hline
    \end{tabular}
    \caption{Criteria used when determining the photometricity of a given observing night. Photometric nights have stable atmospheric conditions and negligible cloud extinction, ideal conditions for spectrophotometry. Non-photometric nights have unstable/turbulent atmosphere and/or detectable cloud cover. The intermediate classification, semi-photometric nights, do not exceed any of the non-photometric thresholds but also do not meet all of the photometric criteria.}
    \label{tab:phot}
\end{table*}

SNIFS was designed for transient spectrophotometry \citep{aldering2002} but several steps are necessary to ensure the nightly flux calibration is accurate. We present our atmospheric model in \S\ref{subsec:specphot.atm}, outline the criteria for photometric conditions in \S\ref{subsec:specphot.photometricity}, and describe the procedure for deriving the nightly atmospheric and flux solutions in \S\ref{subsec:specphot.process}.

\subsection{Atmospheric Transmission}\label{subsec:specphot.atm}

\begin{figure}
    \centering
    \includegraphics[width=\linewidth]{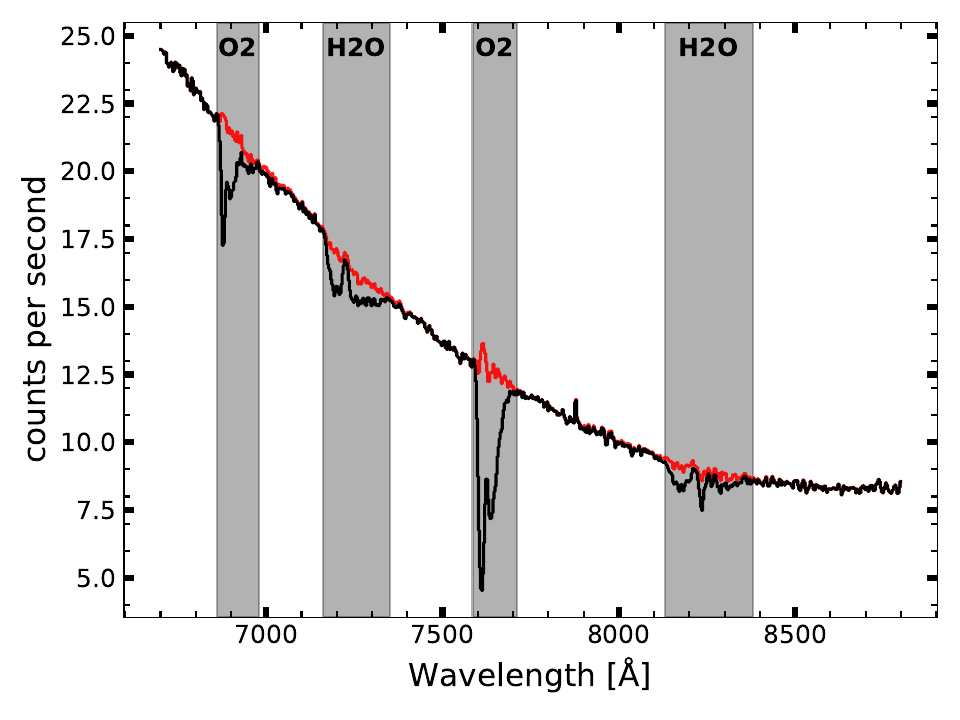}
    \caption{A randomly-selected \stdstar spectrum before (black) and after (red) correcting for telluric absorption.}
    \label{fig:tellcorr}
\end{figure}

Spectra must be corrected for atmospheric extinction in order to be placed on a reliable absolute flux scale. Following \citet{buton2013}, the total atmospheric extinction (excluding telluric lines) is comprised of three components: Rayleigh scattering, aerosol scattering, and ozone absorption. The Rayleigh component is essentially non-variable \citep{buton2013}, leaving three free parameters describing the two remaining components. The ozone template is multiplied by a dimensionless scale factor corresponding to the strength of the absorption features at $<3300~$\AAA and $5000-7000$~\AAA. The aerosol component is parameterized as a power-law with 2 free parameters \citep[e.g., ][]{reimann1992}, 

\begin{equation}\label{eq:aerosols}
    k_{\rm A} (\lambda) = \tau_1 \times (\lambda/1~\mu \rm m)^{-\mathring{a}}
\end{equation}

\noindent with inverse wavelength exponent \r{a} and 1\um optical depth $\tau_1$. 

Gaussian priors on the atmospheric parameters are similar to those adopted by \citet{buton2013}, namely $\mathring{\rm{a}} = 1\pm3$ and $\log_{10}\tau_1 = -2 \pm 1$. The O$_3$ absorption template we generate with \textsc{molecfit} \citep{molecfit1, molecfit2} is similar, but not identical, to the template used by \citet{buton2013}. Our ozone template is normalized such that median conditions on Maunakea produce a scale factor of 1 and we adopt the same (fractional) uncertainty as \citet{buton2013} of $20\%$. 

\subsection{Nightly Photometricity}\label{subsec:specphot.photometricity}

The photometric stability (i.e., photometricity) of each night must be assessed to determine the reliability of the flux solution. Several probes are used in conjunction to determine if a given night is photometric. Table~\ref{tab:phot} provides a summary of the criteria which we elaborate upon here. 

\textit{SkyProbe}: The SkyProbe atmospheric monitor \citep{skyprobe1, skyprobe2, skyprobe3} is part of the Canada-France-Hawai'i Telescope (CFHT) and is dedicated to real-time measurements of the atmospheric transmission. As outlined in \citet{buton2013}, some cleaning of the raw SkyProbe transmission measurements is necessary to eliminate spurious points and correct inter-pointing offsets. After processing, transmission RMS values of $\leq2.5\%$ indicate excellent conditions and $>5\%$ RMS indicate poor conditions. 

\textit{Quality of the PSF fit}: How well the PSF is fit by the model (Eq.~\ref{eq:psf}) is another avenue for constraining the photometricity of a given exposure. Poorly-modeled PSFs can be attributed to several different aspects of the observing and atmospheric conditions. For example, large and/or highly variable seeing make it difficult to precisely measure the PSF shape parameters because of the trade-offs between the Gaussian core, the Moffat wings, and the background sky flux. No exposure with a seeing of $>2\arcsec$ is considered photometric, as \citet{rubin2022} showed the PSF becomes poorly behaved under these conditions. Other factors that can degrade the quality of the PSF fit include imprecise focusing or telescope jitter caused by wind shake. Empirically, reduced chi-squared statistics of $\chi^2/\nu > 1.5$ represent a poor fit to the data and thus unlikely to be photometric and PSF fits with $\chi^2/\nu \le 1.2$ are considered photometric.

\textit{Photometric Zeropoints}: The uncertainty on the derived photometric zero points constrains the stability of the atmosphere over the course of the night, as non-photometric conditions will introduce intrinsic scatter into the nightly photometry. We use different thresholds for the $V$-band acquisition images and the MF filters because there is typically an order of magnitude or more stars available for the $V$-band observations. Therefore, the $V$-band solution has typical uncertainties $\lesssim 0.01$~mag whereas the MF zeropoint uncertainties are typically $\sim 0.02-0.06$~mag. 

\textit{Flux solution}: The RMS of the derived flux calibration is a measure of how well the observed \stdstar spectra are reproduced by our atmospheric and instrumental solutions. SNIFS has a single-exposure extraction floor of a few percent \citep{buton2009, rubin2022} so we set an upper limit on the flux calibration RMS of 3\% for photometric conditions. The non-photometric regime is characterized by \stdstar RMS in excess of 5\%, our desired spectrophotometric accuracy.

\textit{Photometric exposure fraction}: The photometric exposure fraction is determined by comparing the number of exposures that meet the single-exposure photometric criteria, namely the RMS of the guide-star flux and the measured seeing, to the total number of exposures obtained that night.  All exposures must be photometric for a given night to be considered photometric and nights with $<75\%$ of exposures meeting the photometric criteria are considered non-photometric because the aerosol component of the atmospheric model can compensate for persistent low-extinction cloud cover (c.f. Fig. 13 from \citealp{buton2013}). Bright ($V \approx 6$~mag) \stdstar observations are excluded when calculating the photometric exposure fraction because the short exposure times ($\approx1$~s) can produce aberrant PSF shapes even under good conditions. 

Combining the criteria outlined above and in Table~\ref{tab:phot}, we can assess the photometricity of a given night. The photometric criteria are restrictive by design so all photometric criteria must be met for a given night to be considered photometric. Similarly, a night meeting or exceeding any of the non-photometric criteria are deemed non-photometric. The intermediate classification, semi-photometric nights, occurs when none of the non-photometric criteria are met, but neither are all of the photometric criteria. These nights usually are affected by passing and/or thin clouds that only affect some of the exposures. For these nights, photometricity is handled on a ``rolling'' basis where the longest uninterrupted time-span of exposures meeting the photometric criteria is considered photometric, and the remaining time is non-photometric. This relies mostly on the photometric exposure fraction, as no more than 25\% of the exposures can be considered non-photometric to ensure that a reliable atmospheric and flux solution can still be derived. 

Note that the concurrent $V$-band or MF images provide spectrophotometric capabilities even under non-photometric conditions but with a significantly higher uncertainty ($\sim 10-20\%$) compared to photometric conditions ($\leq 5\%$) due to the limited number of calibration stars in the P-channel FoV. A full description and analysis of the long-term spectrophotometric accuracy, including uncertainty estimation for non-photometric conditions, will be presented in a forthcoming paper.

\subsection{Deriving the nightly atmospheric and flux solution}\label{subsec:specphot.process}

\begin{figure}
    \centering
    \includegraphics[width=\linewidth]{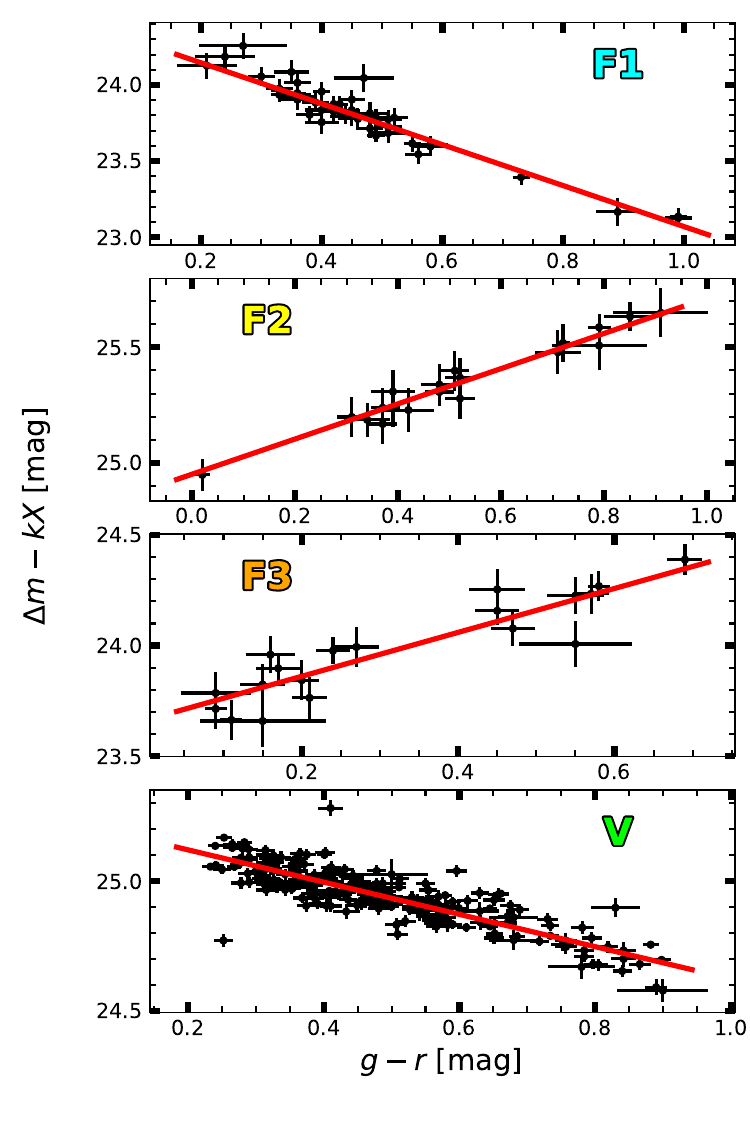}
    \caption{Results from fitting the photometric coefficients (Eq.~\ref{eq:photcal}) to photometry from the $V$-band acquisition images and the $F1$, $F2$, and $F3$ filters. The fitted coefficients for each filter provided in Table~\ref{tab:phottbl} constrain the atmospheric transmission (see \S\ref{subsec:specphot.atm}) and the nightly photometricity (see \S\ref{subsec:specphot.photometricity}). Note that airmass corrections have been applied.}
    \label{fig:photfits}
\end{figure}

\begin{figure*}
    \centering
    \includegraphics[width=\linewidth]{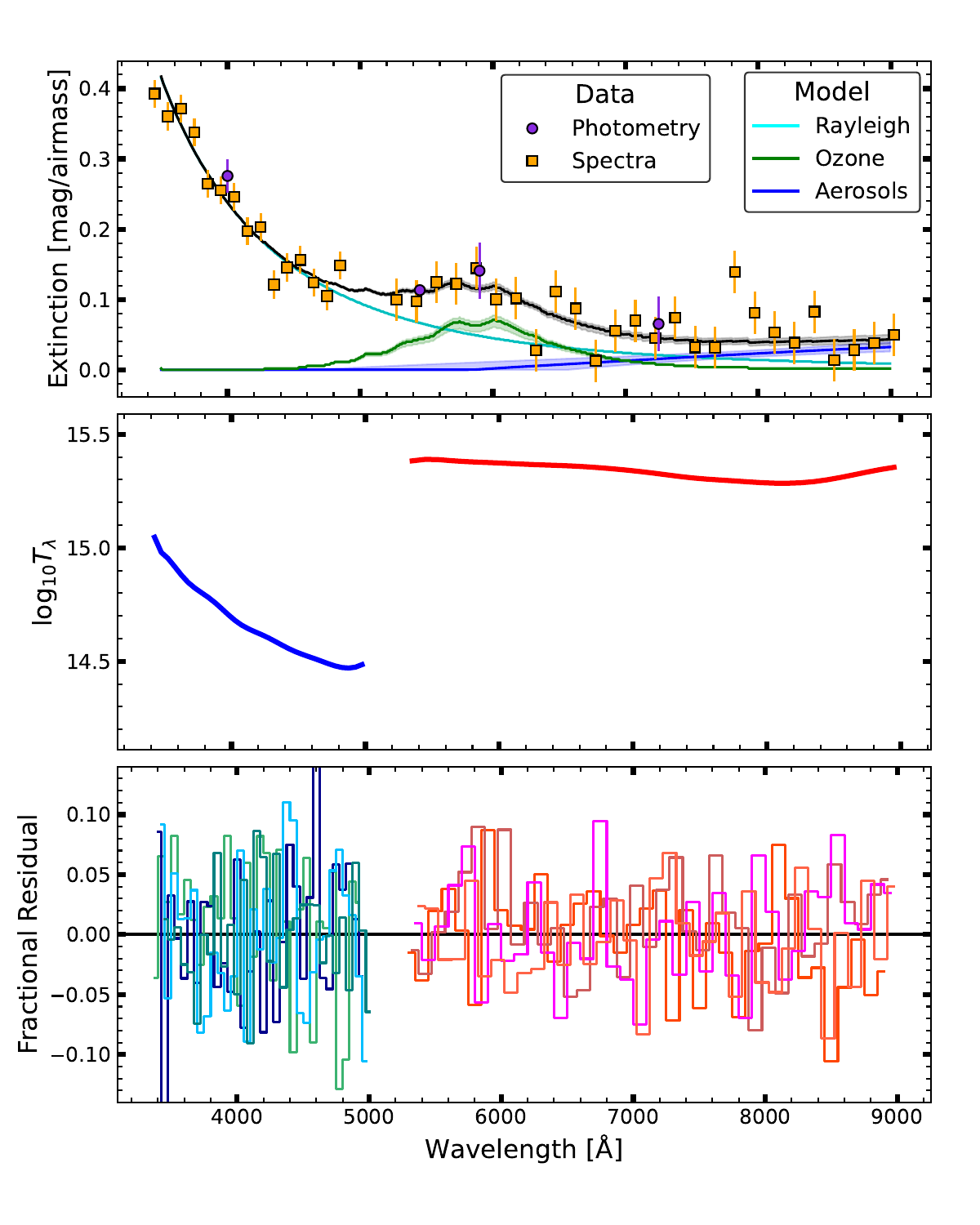}
    \caption{Example atmospheric and flux calibration solution for a typical night, UT 2020-02-15. \textit{Top}: Atmospheric extinction curve derived by jointly fitting the photometry (purple circles, see Fig.~\ref{fig:photfits}) and the \stdstar spectra (orange squares). \textit{Middle}: Derived flux-calibration curve for the night. Note that the normalized spectral flat field has been applied. \textit{Bottom}: Per-spectrum residuals from our flux calibration pipeline for the 4 \stdstar spectra. Individual spectra are arbitrarily offset along the x-axis for visual clarity. }
    \label{fig:atm_resp}
\end{figure*}

\begin{table}
    \centering
    \begin{tabular}{lccc}
    \hline\hline
    Filter & $z$ & $t$ & $k$ \\
     & [mag] & & [mag/airmass] \\\hline
    $F1$ & $24.42\pm0.03$ & $-1.35\pm0.05$ & $0.27\pm0.02$ \\
    $F2$ & $24.95\pm0.04$ & $0.76\pm0.03$ & $0.14\pm0.04$\\ 
    $F3$ & $23.66\pm0.05$ & $0.99\pm0.07$ & $0.06\pm0.04$\\
    $V$ & $25.25\pm0.01$ & $-0.62\pm0.01$ & $0.11\pm0.01$  \\
    \hline\hline
    \end{tabular}
    \caption{Filter coefficients from Eq. \ref{eq:photcal} for the data shown in Fig. \ref{fig:photfits} and the top panel of Fig \ref{fig:atm_resp}.} 
    \label{tab:phottbl}
\end{table}

We utilize both photometry and spectroscopy to derive each night's atmospheric parameters. First, we leverage the P-channel photometry to derive broadband wavelength-dependent extinction coefficients. We typically obtain $\geq 15$ $V$-band acquisition images and $5-7$ MF images during a typical observing night. The equation

\begin{equation}\label{eq:photcal}
    m_{\rm{ref}} - m_{\rm{obs}} = \Delta m = z + t(g-r) + kX
\end{equation}

\noindent converts the observed instrumental magnitude, $m_{\rm{obs}}$, into the reference filter magnitude $m_{\rm{ref}}$ at airmass $X$ using the zeropoint $z$, atmospheric extinction coefficient $k$, and color coefficient $t$ corresponding to the $g-r$ color from \textsc{refcat} \citep{refcatref}. Only the $F1$ ($\lambda_{\rm{eff}}\approx 4000~$\AA), $F2$ ($\lambda_{\rm{eff}} \approx 5900~$\AA), and $F3$ ($\lambda_{\rm{eff}} \approx 7250~$\AA) filters are used from the MF mosaic due to the redder $F4$ and $F5$ filters being contaminated by the strong H$_2$O absorption from $\sim 0.9-1~\mu\rm m$. Fig.~\ref{fig:photfits} shows example results from fitting Eq.~\ref{eq:photcal} to each set of filter photometry with the coefficients provided in Table~\ref{tab:phottbl}. 

Then, the observed spectra are corrected for molecular absorption bands caused by O$_2$ and H$_2$O. We start with a high-resolution absorption template generated using \textsc{molecfit} \citep{molecfit1, molecfit2}. These absorption bands are typically saturated and follow an airmass dependence of $k_\oplus =  sX^\rho$, with $\rho$ representing the saturation parameter, $X$ is the exposure's airmass, and $s$ is a dimensionless scale factor. We adopt a non-variable $\rho_{\rm{O}2} = 0.6$ as this has been shown to produce excellent results for Maunakea \citep{buton2013}. The H$_2$O saturation parameter $\rho_{\rm{H2O}}$ is also fixed to 0.6 if fewer than 5 \stdstar observations are available, otherwise $\rho_{\rm{H2O}}$ is allowed to vary. We include an additional free parameter that convolves the high-resolution absorption template to account for minor wavelength differences between the spectra and the \textsc{molecfit} model. An example telluric correction is shown in Fig.~\ref{fig:tellcorr}. We do not attempt to correct for the H$_2$O absorption band at $>9000$\AAA because the continuum level is uncertain. 

Finally, we use the telluric-corrected \stdstar spectra to jointly fit the atmospheric extinction and the flux calibration solution. Table 1 in \citet{rubin2022} lists the spectrophotometric \stdstar targets observed with SNIFS. After binning the spectra into ``meta-slices'' (see \S\ref{subsec:pipeline.BR}), we fit

\begin{equation}\label{eq:fluxcal}
    S_\lambda = \bar{S_\lambda} \times k_\lambda(X) \times T_\lambda
\end{equation}

\noindent with observed spectrum $S_\lambda$, the corresponding reference spectrum $\bar{S_\lambda}$, total atmospheric extinction $k_\lambda$ at wavelength $\lambda$, and the combined instrumental and telescopic response $T_\lambda$ modeled using a low-order spline. \citet{buton2013} show that $T_\lambda$ is a smoothly-varying function of wavelength on scales less than the meta-slice width ($100$~\AAA for the B-channel, $150$~\AAA for the R-channel).

On photometric nights, the atmosphere is considered non-variable and cloud extinction is negligible, allowing direct calculation of $k_\lambda$ and $T_\lambda$ using the observed \stdstar spectra. However, a different approach must be adopted for non-photometric conditions where thin and/or intermittent clouds can diminish the observed flux. On these nights, each \stdstar observation is allowed to have non-zero cloud cover, parameterized as a grey (wavelength-independent) extinction component\footnote{\citet{buton2013} showed that the chromaticity of cloud extinction is $<1\%$. }, that diminishes the observed flux by a factor of $f$. We require $0 < f < 1$ and use the corresponding MF or $V$-band photometry, if available, to place an additional prior on cloud cover during the exposure with a typical uncertainty of $\delta f / f \sim 10\%$. If $<3$ \stdstar observations are obtained for a given night, usually due to unfavorable weather conditions, we adopt the mean atmospheric extinction from \citet{buton2013}.

Example results of the atmospheric and flux calibration routines are shown for a typical photometric night, UT 2020-02-15, where 4 \stdstar observations were obtained. The telluric correction for a randomly-selected \stdstar spectrum is shown in Fig.~\ref{fig:tellcorr}, results from fitting Eq.~\ref{eq:photcal} to the photometry are provided in Fig.~\ref{fig:photfits}, and Fig.~\ref{fig:atm_resp} shows the derived atmospheric extinction curve, including its decomposition into the various physical components, and the derived flux calibration solution. The bottom panel of Fig.~\ref{fig:atm_resp} shows the per-spectrum B-channel and R-channel residuals.

\section{Initial Results}\label{sec:conclusion}

\begin{figure}
    \centering
    \includegraphics[width=\linewidth]{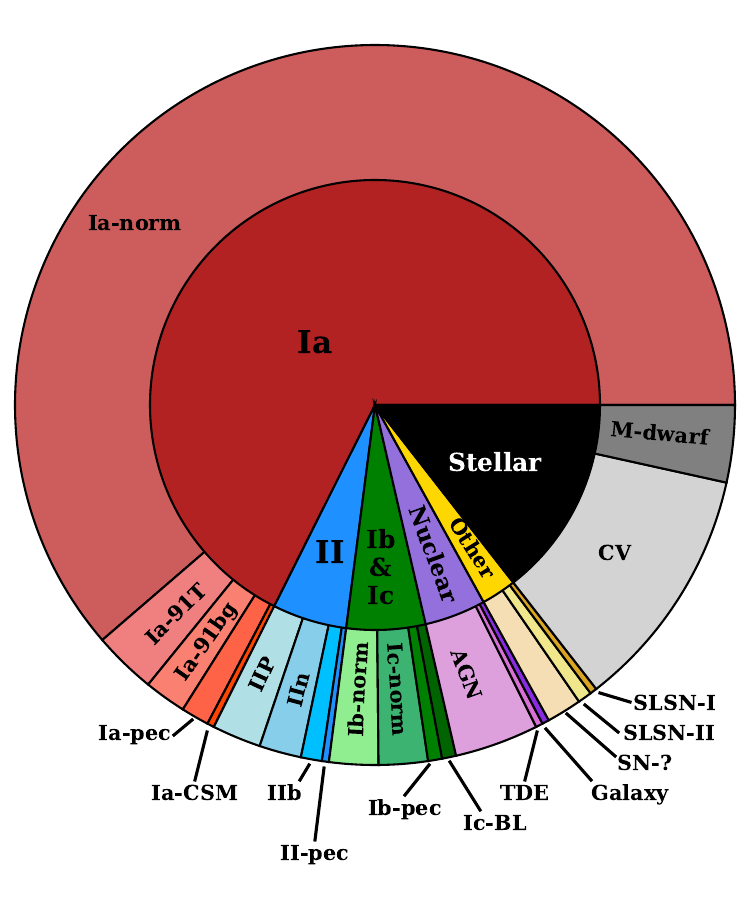}
    \caption{Distribution of transient classification types from the first 3 years of the SCAT survey.}
    \label{fig:classpie}
\end{figure}

Over the course of the 2018B$-$2020B observing semesters (i.e., UT 2018-02-01 through 2021-01-31), we obtained 635 science exposures over 101 nights\footnote{Excluding nights where no observations were taken due to weather, maintenance, or COVID-related closures. }. Roughly 75\% of the observations were dedicated to classifying transients and the remaining $\sim 25\%$ are follow-up observations. We provide details on classifications and follow-up observations in \S\ref{subsec:discuss.classify} and \S\ref{subsec:discuss.followup}, respectively. \S\ref{subsec:discuss.data} describes our data policy, and future plans for the SCAT survey are detailed in \S\ref{subsec:discuss.future}.

\subsection{Transient Classification}\label{subsec:discuss.classify}

The classification spectra are extracted with a quick-reduction pipeline which is very similar to the one described in \S\ref{sec:pipeline} with two notable differences: 1) spectra are extracted using aperture photometry instead of PSF fitting, 2) the flux solution is relative instead of absolute, and 3) no effort is made to correct the dichroic crossover region. Extracting the spectra using aperture photometry instead of PSF fitting expedites the reduction process, allowing us to classify objects during the course of a night, but incurs drawbacks such as poorer host-galaxy subtraction. Similarly, adopting a relative flux solution allows the use of observations from other observing nights to construct the flux solution and avoid deriving a nightly atmospheric solution. Thus, classification spectra released via TNS are not guaranteed to have reliable absolute flux calibration and should not be used for such purposes without proper vetting. The dichroic, which deflects light between the B and R spectroscopic channels, has humidity-dependent transmission due to the use of a water-based coating. Special spectroscopic flat field images are taken before each science exposure to mitigate dichroic transmission variations but these spectral flats are not currently applied by the quick-reduction pipeline. Thus, the $\approx 5000-5200$~\AAA region should be treated with caution when analyzing classification spectra.

Transient spectra are classified via visual inspection with assistance from the SuperNova Identification (SNID) software \citep{blondin2007}. Fig.~\ref{fig:classpie} shows the different types of transients classified during the first 3 years of SCAT. 482 of the 635 exposures ($75.9\%$) were obtained for classification purposes, of which 369\footnote{The best way to reproduce our numbers through the TNS web query is to use the following input values: `Discovery Date Range' = (2018-02-01, 2021-01-31), Classification Instrument = `UH88 - SNIFS', and check the `Official Classification' box (\href{https://www.wis-tns.org/search?&discovered_period_value=&discovered_period_units=months&unclassified_at=0&classified_sne=0&include_frb=0&name=&name_like=0&isTNS_AT=all&public=all&ra=&decl=&radius=&coords_unit=arcsec&reporting_groupid[]=null&groupid[]=null&classifier_groupid[]=null&objtype[]=null&at_type[]=null&date_start[date]=2018-02-01&date_end[date]=2021-01-31&discovery_mag_min=&discovery_mag_max=&internal_name=&discoverer=&classifier=&spectra_count=&redshift_min=&redshift_max=&hostname=&ext_catid=&ra_range_min=&ra_range_max=&decl_range_min=&decl_range_max=&discovery_instrument[]=null&classification_instrument[]=11&associated_groups[]=null&official_discovery=0&official_classification=1&at_rep_remarks=&class_rep_remarks=&frb_repeat=all&frb_repeater_of_objid=&frb_measured_redshift=0&frb_dm_range_min=&frb_dm_range_max=&frb_rm_range_min=&frb_rm_range_max=&frb_snr_range_min=&frb_snr_range_max=&frb_flux_range_min=&frb_flux_range_max=&num_page=50&display[redshift]=1&display[hostname]=1&display[host_redshift]=1&display[source_group_name]=1&display[classifying_source_group_name]=1&display[discovering_instrument_name]=0&display[classifing_instrument_name]=0&display[programs_name]=0&display[internal_name]=1&display[isTNS_AT]=0&display[public]=1&display[end_pop_period]=0&display[spectra_count]=1&display[discoverymag]=1&display[discmagfilter]=1&display[discoverydate]=1&display[discoverer]=1&display[remarks]=0&display[sources]=0&display[bibcode]=0&display[ext_catalogs]=0&format=html&edit[type]=&edit[objname]=&edit[id]=&sort=asc&order=classifying_source_group_name}{link}).} ($\approx 75\%$) were classified and publicly released through TNS. Another $\approx 15\%$ of the classification targets were first classified by another transient classification program such as ePESSTO+ \citep{smartt2015} or the ZTF BTS \citep{fremling2020, perley2020}. The remaining $\sim 10\%$ of the exposures had a too-low signal for conclusive typing, usually obtained during unfavorable weather conditions.

\subsection{Follow-Up Targets}\label{subsec:discuss.followup}

The remaining $\approx 25\%$ of our observing time is reserved for extensive follow-up observations of interesting and unique discoveries from the world's sky surveys. Each follow-up target is typically observed several times over the course of the follow-up campaign which can span one or multiple semesters. While some of the follow-up spectra are still being analyzed, several papers have already been published using SCAT spectra on a variety of transient phenomena.

Many of the papers published using SCAT observations focus on TDEs. We classified PS18kh as a TDE during our first observing night in March 2018 \citep{18kh_ATel} and initiated an extensive follow-up campaign leading to the first discovery of a TDE with an asymmetric accretion disk \citep{holoien2019}. Other TDEs with SCAT spectra include ASASSN-19dj \citep{hinkle2020}, ASASSN-20hx \citep{hinkle2021}, and ATLAS18mlw \citep{hinkle2022}. It is unclear if ASASSN-18jd is a TDE or related to AGN activity \citep{neustadt2020}, part of the growing number of Ambiguous Nuclear Transients (ANTs; e.g., \citealp{ants, hinkle2023}). 

Additionally, we have observed several other transients involving the interplay between black holes and their accretion environments including the low-mass X-ray binary (LMXB) ASASSN-18ey/MAXI J1820+070 \citep{tucker2018} and the changing-look blazar B2 1402+32 \citep{mishra2021}. Finally, we have also observed the young stellar object (YSO) outburst Gaia 19bey \citep{hodapp2020} and the intriguing SN Icn 2021csp \citep{fraser2021}, part of the emerging class of SNe Ic with narrow C/O emission lines \citep[e.g., ][]{galyam2022, pellegrino2022}. Many more papers are in various stages of analysis and publication, and this number will begin to increase dramatically due to our partnerships with the Hawai'i Supernova Flows (HSF, Do et al., in prep) and the Precision Observations of Infant Supernovae \citep[POISE; ][]{poise} collaborations. 

\subsection{Data Policy and Releases}\label{subsec:discuss.data}

All spectra used to classify transients are made publicly available to the community through the TNS website. However, as noted in \S\ref{subsec:discuss.classify}, the classification spectra are reduced with a simplified version of the reduction pipeline and are not guaranteed to have absolute spectrophotometry. Anyone wishing to obtain spectrophotometric-quality classification spectra should contact the SCAT team.

We plan on periodic data releases for the spectrophotometric observations. The precision spectrophotometry of our SCAT observations will be essential to test and constrain the next generation of physical models for a vast array of transient phenomena. 

\subsection{Future Plans}\label{subsec:discuss.future}

SCAT will benefit from several upgrades to the UH2.2m/SNIFS system over the next few months. The tertiary mirror was replaced in early 2022 which provides an increase in the total system throughput and allows observers to quickly switch between instruments. This capability is especially useful for STACam, a wide-field imager ($14.5\arcmin\times14.5\arcmin$ FoV at $0\farcs08$/pixel) equipped with $grizy$ filters. Rapid switches between SNIFS and STACam provide contemporaneous imaging capabilities during the readout of the SNIFS CCDs. Thus, we can obtain improved multi-color photometry of transients and additional constraints on the atmospheric extinction (e.g., Figs. \ref{fig:photfits} and \ref{fig:atm_resp}) with no additional observing time required. 

Another instrument recently added to the UH 2.2-m is Robo-AO-2 \citep{baranec2018, baranec2021}, an optical and near-infrared laser adaptive optics imaging system that provides spatial resolution approaching that of the Hubble Space Telescope. While the previous generation Robo-AO system \citep{baranec2014} was typically used for searching for companions to stellar and planetary systems (e.g.,  \citealt{ziegler2018,lamman2020,salama2021}), SCAT will use the improved high-acuity imaging capabilities of Robo-AO-2 to quickly assess if transients close to the center of their host galaxies are truly nuclear, as with AGN variability and TDEs, or non-nuclear with a small projected separation.

Additionally, work is underway to make the UH2.2m system entirely robotic so no human intervention is necessary for nightly observations (Shappee et al., in prep). Observations will be scheduled automatically based on the score assigned by the telescope allocation committee. This will increase the amount of time spent on science targets because calibration data, mainly \stdstar spectra, can be reserved for non-ideal weather conditions or nights with high moon illumination. Robotic operations also open many avenues for improving the per-night efficiency and expanding the capabilities of SCAT for observing rare and/or fast-declining transients. 

Science data from LSST is expected in mid-2024\footnote{\url{https://www.lsst.org/about/project-status}} and subsequently, the number of transients discovered in the southern hemisphere is expected to increase drastically. The UH2.2m, located on Maunakea at a latitude of $\approx 19\deg$, can observe objects with declination $\delta \gtrsim -40~\deg$. As such, there will be significant overlap between UH2.2m's observable sky area and the LSST footprint ($\delta \lesssim +30~\deg$). While most transients discovered by LSST will be too faint for classification with SNIFS, SCAT will observe some of these discoveries as they rise to peak brightness and allow larger-aperture telescopes to focus their time on fainter targets. 

Starting in 2023, SCAT will also have access to the Wide-Field Spectrograph \citep[WiFeS; ][]{dopita2007} on the Australian National University 2.3-meter (ANU2.3m) telescope. There are many similarities between WiFeS+ANU2.3m and SNIFS+UH2.2m. Both are integral-field spectrographs on moderate-aperture telescopes covering the full optical wavelength range. WiFeS has a larger FoV ($25\arcsec\times38\arcsec$) and slightly higher spectral resolution ($R\sim 3000$). Incorporating WiFeS+ANU2.3m will improve our southern hemisphere capabilities where transients are already discovered by the ASAS-SN and ATLAS sky surveys, in addition to the upcoming LSST. Thus, SCAT will be able to obtain spectrophotometry of any $\lesssim 19$~mag transient regardless of sky position or discovery survey.

Multi-messenger synergy is an exciting avenue for future expansion, especially considering the upcoming improvements to SNIFS+UH2.2m and the inclusion of WiFeS+ANU2.3m. The astronomical community still lacks an understanding of which events detected via gravitational waves (e.g., LIGO/VIRGO, \citealp{aasi2015, acernese2015, abbott2021}) or neutrinos (e.g., IceCube, \citealp{aartsen2013}) will have optical counterparts. The sky localizations of these events usually span $\sim10^{2}-10^3$ square degrees \citep{abbott2020} leading to many potential optical counterparts \citep[e.g., ][]{coughlin2020, dejaeger2022, necker2022}. These counterparts require spectroscopic confirmation to separate the multi-messenger sources from typical optical transients such as SNe. However, early identification and follow-up of multi-messenger sources provides unprecedented insight into the physical processes governing these extreme events \citep[e.g., ][]{coulter2017, shappee2017, valenti2017, aartsen2018, garrappa2019}. Robotic triggering of SNIFS will provide rapid follow-up capabilities for observing the extreme events in our Universe. Additionally, SNIFS readily provides precise spectrophotometric calibration whereas target-of-opportunity observations rarely contain the necessary calibrations for absolute spectrophotometry.

The proliferation of sky surveys has made it an exciting time for studying and understanding astrophysical transients. SCAT is specifically designed to capitalize on this discovery deluge and will provide the community with precise spectrophotometry necessary for testing the next generation of theoretical models. 

\section*{Software}
numpy \citep{numpy}, astropy \citep{astropy}, lmfit \citep{lmfit}, emcee \citep{emcee}, spectres \citep{spectres}, matplotlib \citep{matplotlib}, photutils \citep{photutils}, scipy \citep{scipy}, pandas \citep{pandas}, astroscrappy \citep{astroscrappy}

\section*{Facilities}
\facility{UH: 2.2m}

\section*{Acknowledgements}

M.A.T. acknowledges support from the DOE CSGF through grant DE-SC0019323. This material is based upon work supported by the National Science Foundation Graduate Research Fellowship Program under Grant No. 2236415. G.A acknowledges support from the Director, Office of Science, Office of High Energy Physics, of the U.S. Department of Energy under Contract No.~DE-AC02-05CH11231. Any opinions, findings, and conclusions or recommendations expressed in this material are those of the author(s) and do not necessarily reflect the views of the National Science Foundation. UH2.2m robotization: NSF grant AST-1920392. Robo-AO-2: NSF grant AST-1712014.

Pan-STARRS is a project of the Institute for Astronomy of the University of Hawaii, and is supported by the NASA SSO Near Earth Observation Program under grants 80NSSC18K0971, NNX14AM74G, NNX12AR65G, NNX13AQ47G, NNX08AR22G, 80NSSC21K1572  and by the State of Hawaii.

\bibliography{ref}{}
\bibliographystyle{aasjournal}



\end{document}